\documentclass[journal]{IEEEtran}
\newcommand{\revised}[1]{{\color{black}#1}}
\usepackage{etoolbox}
\usepackage{array}
\usepackage{tipa}

\makeatletter
\patchcmd{\@twocolumnfalse}{\setpagewiselinenumbers}{\setpagewiselinenumbers\linenumbers}
{\setpagewiselinenumbers\linenumbers}{}
\makeatother

\usepackage{amsmath}
\usepackage{algorithm}
\usepackage{algpseudocode}
\usepackage{booktabs}
\usepackage{multirow}
\usepackage[utf8]{inputenc} % allow utf-8 input
\usepackage[T1]{fontenc}    % use 8-bit T1 fonts
\usepackage{hyperref}       % hyperlinks
\usepackage{url}            % simple URL typesetting
\usepackage{booktabs}       % professional-quality tables
\usepackage{amsfonts}       % blackboard math symbols
\usepackage{nicefrac}       % compact symbols for 1/2, etc.
\usepackage{microtype}      % microtypography
\usepackage{xcolor}         % colors
\usepackage{graphicx}
\usepackage{float}
\usepackage[numbers, sort&compress]{natbib}

\newfloat{figtab}{htb}{fgtb}
\makeatletter
  \newcommand\figcaption{\def\@captype{figure}\caption}
  \newcommand\tabcaption{\def\@captype{table}\caption}
\makeatother

\def\BibTeX{{\rm B\kern-.05em{\sc i\kern-.025em b}\kern-.08em T\kern-.1667em\lower.7ex\HbOx{E}\kern-.125emX}}

\usepackage{lineno}
\usepackage{etoolbox}
\makeatletter
\patchcmd{\@twocolumnfalse}{\setpagewiselinenumbers}{\setpagewiselinenumbers\linenumbers}
{\setpagewiselinenumbers\linenumbers}{}
\makeatother

\title{Neural Spelling: A Spell-Based BCI System for Language Neural Decoding}
% ...
\author{
\IEEEauthorblockN{Xiaowei Jiang\textsuperscript{1}\textsuperscript{\dag}, Jinzhao Zhou\textsuperscript{1}\textsuperscript{\dag}, Yiqun Duan\textsuperscript{1}, Ziyi Zhao\textsuperscript{1}, Yu-Cheng Chang\textsuperscript{1}, Thomas Do\textsuperscript{1}, Chin-Teng Lin\textsuperscript{*}\textsuperscript{1}}\\
\IEEEauthorblockA{\textsuperscript{1}Computational Intelligence and Brain Computer Interface Lab, School of Computer Science, \\ Faculty of Engineering and Information Technology, University of Technology Sydney}
\and
\thanks{\textsuperscript{\dag}Xiaowei Jiang and Jinzhao Zhou contributed equally to this work.}
\thanks{\textsuperscript{*}Corresponding author: Xiaowei Jiang. Email: xiaowei.jiang-1@student.uts.edu.au}
}

\makeatletter
\def\bstctlcite{\@ifnextchar[{\@bstctlcite}{\@bstctlcite[@auxout]}}
\def\@bstctlcite[#1]#2{\@bsphack
  \@for\@citeb:=#2\do{%
    \edef\@citeb{\expandafter\@firstofone\@citeb}%
    \if@filesw\immediate\write\csname #1\endcsname{\string\citation{\@citeb}}\fi}%
  \@esphack}
\makeatother

\begin{document}
% \linenumbers
\maketitle
\begin{abstract}

\textbf{Objective:} Brain–computer interfaces (BCIs) can \revised{support} communication by translating neural activity into text, yet existing non-invasive systems rarely cover the full alphabet, limiting their practical use.
\textbf{Methods:} We propose a novel non-invasive EEG-based BCI framework, Curriculum-based Neural Spelling (CNS), that decodes all 26 English letters by first learning neural patterns associated with handwriting trajectories. A Generative AI (GenAI) module based on large language models (LLMs) is then integrated to refine spell-based neural decoding and correct semantic errors.
\textbf{Results:} The proposed system achieves robust letter-level decoding and accurate sentence reconstruction across users, outperforming conventional EEGNet and hybrid CNN–RNN baselines. GenAI correction further reduces word error rates and enhances decoding fluency.
\textbf{Conclusion:} Combining EEG-based neural spelling with generative language modeling enables full-alphabet decoding and significantly improves linguistic accuracy in non-invasive BCIs.
\textbf{Significance:} This work demonstrates how integrating GenAI with neural decoding
can bridge the gap between noisy signal-level predictions and coherent language-level
outputs, \revised{establishing a system-level framework for full-alphabet neural
spelling and adaptive language-level correction under non-invasive EEG constraints.}

\end{abstract}


\bstctlcite{BSTcontrol}

\section{Introduction}

\IEEEPARstart{B}{rain-computer} interfaces (BCIs) have emerged as a pivotal area of research within human-computer interaction (HCI), distinguished by their capacity to seamlessly integrate neural signals with external systems. Pioneering studies such as those by Guo et al.\cite{guo2022ssvep}, Chen et al.\cite{chen2020combination}, Cao et al.\cite{cao2022building}, Nguyen et al.\cite{nguyen2025edge}, and Lin et al.\cite{lin2020direct} underscore BCIs' role in advancing neuroscience and technology. These interfaces create direct communication pathways that are especially beneficial for individuals with limited speech or motor functions. Language decoding represents a critical domain within BCI research, aimed at deciphering the neural correlates of speech and language processing. This capability not only enhances interaction paradigms but also opens new avenues for communication, offering significant improvements in quality of life for those with severe communicative impairments.

\begin{figure}[t]
    \centering
    \includegraphics[width=1\linewidth]{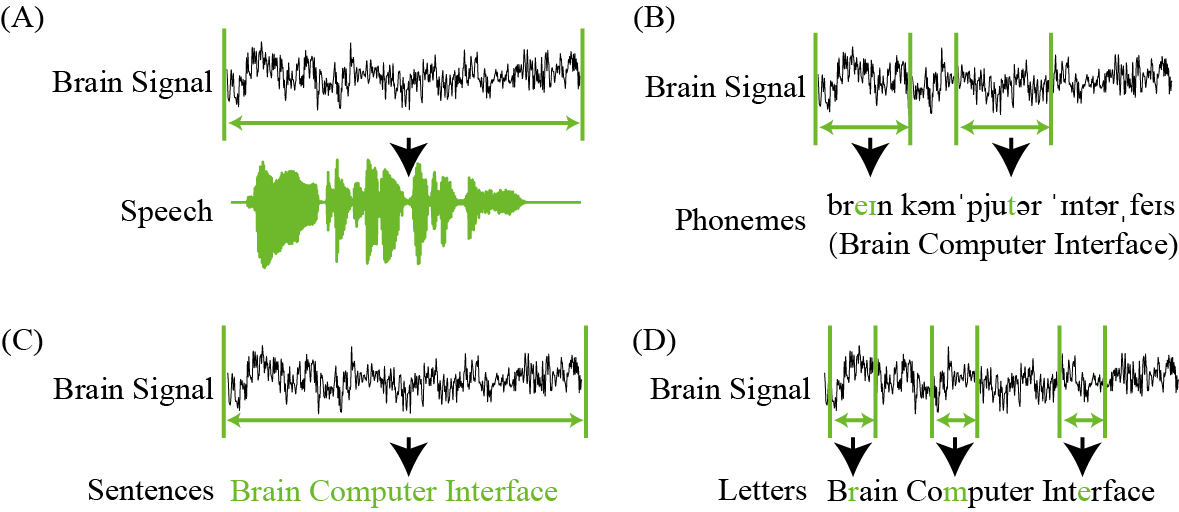}
    \caption{Demonstrations of \textbf{four} typical language decoding frameworks. 
    \textbf{(A)} Direct speech synthesis approach. 
    \textbf{(B)} Spelling-based approach using phonemes.
    \textbf{(C)} Direct text synthesis approach. 
    \textbf{(D)} Spelling-based approach using letters.
    }
    \label{fig:typesintro}
\end{figure}

Initially, BCI research focused on visual-based decoding approaches, such as steady-state visually evoked potentials (SSVEP)\cite{guo2022ssvep}, which, despite their reliability, demand high cognitive effort and are unsuitable for prolonged use\cite{chen2020combination}. These methods often fail to align with natural human language patterns, posing significant usability challenges. The advent of invasive neural decoding technologies marked a significant advancement, allowing for direct interpretation of brain signals via electrocorticography (ECoG) or stereo-EEG (sEEG)\cite{anumanchipalliSpeechSynthesisNeural2019, angrickRealtimeSynthesisImagined2021}. These methods have demonstrated substantial improvements in user performance, significantly enhancing communicative capacities for patients with speech impairments. However, the invasive nature, high cost, and ethical concerns limit their general applicability\cite{willettHighperformanceSpeechNeuroprosthesis2023}. In contrast, non-invasive BCIs, predominantly utilizing electroencephalography (EEG), offer a more accessible alternative. These systems are less obtrusive and more cost-effective, broadening potential user demographics~\cite{tangSemanticReconstructionContinuous2023a, hesam2020evaluation}. Despite the challenges of signal noise and the extensive training required for users, recent studies have demonstrated EEG's potential in effective language decoding~\cite{defossezDecodingSpeechPerception2023b, wangOpenVocabularyElectroencephalographytoText2022a}.

With the rise of Generative AI (GenAI), the integration of large language models (LLMs) into BCI research has opened new avenues for enhancing language decoding~\cite{Metzger2022Generalizable}. This integration can be implemented within two distinct frameworks: directly synthesizing speech or text from brain signals using pre-trained LLMs, as illustrated in Fig.~\ref{fig:typesintro}(A) and Fig.~\ref{fig:typesintro}(C) respectively; and decoding brain signals into minimal language units such as phonemes, as shown in Fig.~\ref{fig:typesintro}(B), or letters, as depicted in Fig.~\ref{fig:typesintro}(D), followed by text generation through natural language processing (NLP) models. 

\begin{figure*}[tp]
    \centering
    \includegraphics[width=1\linewidth]{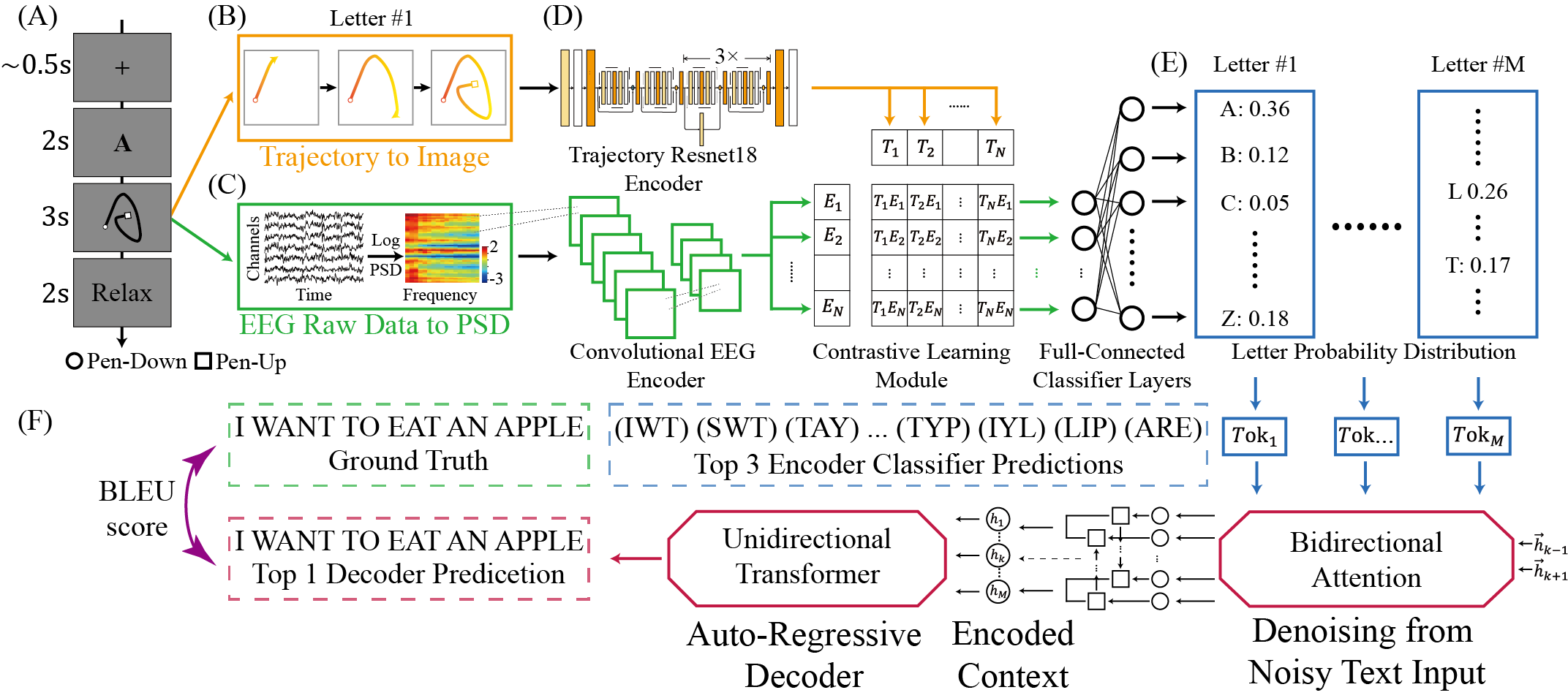}
    \caption{
    \revised{
    \textbf{(A):} The experimental design.
    \textbf{(B):} The trajectory completion over time.
    \textbf{(C):} The power spectral density (PSD) features computed from raw EEG signals.
    \textbf{(D):} The structures of the Trajectory ResNet-18 encoder and the Convolutional EEG encoder.
    \textbf{(E):} The letter probability distribution at the classifier output.
    \textbf{(F):} The structure of the sentence generator.}
    }
    \label{fig:model_arch}
\end{figure*}

The first framework often faces limitations due to the disparity in data scales between brain data and text/image datasets, which can restrict the system to operating within predefined datasets, merely retrieving sentences rather than generating novel content. The second approach, known as the spell-based method \cite{chen2015high, willettHighperformanceBraintotextCommunication2021}, operates by initially decoding neural signals into their minimal representational units, such as the 26 letters, or even nine-key input (T9)~\cite{zhang2018efficient}. It subsequently employs GenAI to construct coherent text in a second stage. This technique is advantageous as it requires fewer categories for the neural decoding model, thereby reducing both the cognitive load on the user and the difficulty associated with model training. A significant benefit of this method is its utilization of extensive textual datasets in training GenAI, particularly for generative error correction (GEC) \cite{yang2024large}, which enhances the accuracy and flexibility of the generated text. Moreover, this method is not constrained by the size of the neural training samples and can adapt swiftly to new linguistic content through fine-tuning of the GenAI, circumventing the need for extensive retraining on new brain data.

The spell-based methods initially utilize N-gram models to calculate the probability of sequential letter occurrences. For instance, the ECoG study aimed at spelling imaged handwritten letters~\cite{willettHighperformanceBraintotextCommunication2021}, and the sEEG study focused on spelling the Pinyin representations of Chinese pronunciations~\cite{Feng2023.11.05.562313}. Subsequently, GenAI technologies have been employed to enhance performance, as demonstrated in studies like~\cite{willettHighperformanceSpeechNeuroprosthesis2023} and~\cite{Metzger2022Generalizable}. These vocal-based methods underscore the potential of GenAI for neural language decoding. However, these approaches have limitations. For example, vocal-based methods require accurate pronunciation by subjects. Given the diversity in pronunciation—such as the thousands of syllables and 44 phonemes in English, including closely related sounds like \textipa{/\textscripta/} and \textipa{/ae/}, along with variations between short and long vowels—it becomes challenging to accurately generate complete sentences. Simultaneously, Mandarin Chinese combines 21 consonant phonemes, 7 vowel phonemes, and 4 tones to create over 400 distinct syllables~\cite{Feng2023.11.05.562313}. This vast array of phonetic elements significantly complicates speech synthesis in both languages. The complexity is further exacerbated in non-invasive speech synthesis applications due to the signal-to-noise ratio (SNR) and the limitations of the training datasets available. These factors together pose substantial challenges in developing robust and accurate speech synthesis systems. Therefore, spelling letters, as opposed to using vocal-based features, presents a more effective option for language decoding. The classification of just 26 letters not only simplifies the process compared to pronunciation-based methods but also enhances performance and applicability, even facilitating cross-linguistic utility, such as seamlessly integrating English and Pinyin. 
\revised{Unlike prior EEG-based handwriting paradigms that typically focus on limited symbol sets or isolated character recognition, the present study systematically investigates full-alphabet (26-letter) decoding and further integrates curriculum-adapted language modeling to enable sentence-level reconstruction.}
To optimize letter input, innovative scenarios have been designed, such as 'air-writing'~\cite{tripathiNeuroAiRDeepLearning2024} and 'paper-writing'~\cite{peiOnlineRecognitionHandwritten2021}.

\revised{
Moreover, cVEP-based systems have demonstrated efficient and reliable performance under non-invasive EEG settings~\cite{10931291}, and online P300 spellers have been established as practical communication systems since the seminal work of Farwell and Donchin~\cite{Farwell1988MentalProsthesis}. More recently, complete sentence-level online spelling pipelines with calibration procedures and large language model integration have also been reported~\cite{Hong2025ChatBCI4ALS}. These approaches represent well-established non-invasive BCI paradigms.

However, a common characteristic of these systems is that they are predominantly \emph{stimulus-driven} or \emph{reactive} in nature, relying on externally evoked neural responses (e.g., visual flickers or oddball paradigms) and predefined symbol sets. While such paradigms achieve high information transfer rates and robust online performance, they are primarily designed to optimize communication efficiency, but this comes at the cost of increased visual fatigue.

In contrast, handwriting-based BCIs aim to decode \emph{active}, self-initiated motor-related neural activity associated with voluntary letter production, enabling more naturalistic and flexible letter generation without reliance on continuous external stimulation. Despite extensive research on non-invasive spellers, whether an active, handwriting-based BCI can reliably decode the full alphabet (26 letters) from scalp EEG, and thereby establish a normative neural reference for neural handwriting under realistic non-invasive constraints, remains an underexplored problem. The present work addresses this gap by focusing on neural handwriting as an active input paradigm, rather than on stimulus-driven letter selection or communication throughput alone.
}

This paper contributes to the evolving landscape of brain-computer interfaces (BCIs) by proposing a hybrid approach that merges the accessibility of non-invasive EEG with the advanced linguistic capabilities of GenAI. We introduce a novel curriculum-based neural spelling framework (CNS) for neural language decoding. This framework employs a convolutional neural network (CNN)-based encoder to initially learn individual-specific letter transition patterns, followed by a curriculum-driven LLM to synthesize sentence texts. The results demonstrate significant enhancements in the performance of language decoding tasks. By integrating these advanced technologies, \revised{we establish a normative neural baseline for non-invasive neural handwriting under realistic EEG constraints. We aim to 
% facilitate more natural, efficient, and inclusive communication methods for individuals with varying physical abilities, 
facilitate a system-level investigation of full-alphabet neural spelling, demonstrating how low-SNR EEG decoding can be coupled with curriculum-adapted generative language models to enable language-level error correction,}
heralding a new era in assistive communication technologies. This introduction outlines the transformation of BCI technology from basic visual decoding to sophisticated language synthesis, setting the stage for a detailed discussion on how current technologies can be integrated and enhanced to better meet the communicative needs of a diverse user population.

The contributions of this work are threefold:
\begin{itemize}
    \item This study is the first to collect and analyze brain dynamics patterns related to EEG-based handwriting of all 26 letters.
    \item We employ a CNN-based model for EEG encoding, which achieves exemplary top-k accuracy, averaging across all subjects.
    \item We propose a novel curriculum supervised fine-tuning method that enables an LLM to learn subject-specific letter transition patterns effectively, thereby enabling high-quality sentence synthesis. This approach has shown promising results, achieving high scores on established metrics.
\end{itemize}

\section{Related Work}

\subsection{Language Decoding}

Recent advances in Language Neural Decoding have significantly enhanced BCI systems. RNN-based models have achieved a $23.8\%$ word error rate on a 125,000-word vocabulary using chronic ECoG signals, highlighting their potential as speech neuroprostheses~\cite{willettHighperformanceSpeechNeuroprosthesis2023}. ECoG-based Speech BCIs have demonstrated stable control of assistive devices for up to three months with minimal calibration, supporting daily unassisted use~\cite{luoStableDecodingSpeech2023}. Additionally, speech synthesis from ECoG signals enables decoding spoken sentences~\cite{anumanchipalliSpeechSynthesisNeural2019}, while contrastive learning models have decoded perceived speech from non-invasive recordings like EEG and MEG, enriching decoding techniques~\cite{defossezDecodingSpeechPerception2023b}. To better predict the text, a novel task, named Cross-Modal Cloze (CMC) task~\cite{zouCrossModalClozeTask2022}, which is to predict the target word with a context as prompt, are proposed, achieving $28.91\%$ accuracy.

Spelling-based BCIs have also progressed, with high-speed systems like the JFPM-based SSVEP speller~\cite{chen2015high} and NeuroAiR, which uses ICA and EEGNet to achieve 44.04$\%$ accuracy in recognizing airwritten letters~\cite{tripathiNeuroAiRDeepLearning2024}. Efforts in tonal language decoding and synthesis~\cite{liuDecodingSynthesizingTonal2023}, handwriting tasks~\cite{willettHighperformanceBraintotextCommunication2021}, and CNN-based classifiers for letter decoding, such as “HELLO, WORLD!”~\cite{peiOnlineRecognitionHandwritten2021}, further highlight BCI versatility.

ECoG systems have provided communication solutions for late-stage ALS patients by decoding English letters~\cite{doi:10.1056/NEJMoa1608085}, while silent spelling mechanisms predict characters in 2.5 seconds, improving practical use~\cite{Metzger2024}. CNNs have decoded sentences~\cite{pmlr-v116-kapur20a} and simple commands~\cite{krishna2019speech, s21206744}, while EEG-based user authentication systems using dynamic signatures have been explored for security~\cite{kumar2019fusion, sainiDonJustSign2018}.

\subsection{Enhancing Text-to-Text Error Correction with Large Language Models}

% Recent advancements in NLP have been largely propelled by the development of LLMs. These models have redefined the paradigm of NLP, moving away from traditional supervised learning approaches toward pretrained models that leverage massive corpora and are fine-tuned or prompted for specific tasks \cite{Min2021RecentAI, Wang2022PreTrainedLM}. This shift has enabled state-of-the-art performance across a wide range of applications.

Advances in NLP have been fueled by the emergence of LLMs, which represent a paradigm shift from traditional supervised learning to pretrained models that leverage vast corpora, subsequently fine-tuned or prompted for specific tasks \cite{Min2021RecentAI, Wang2022PreTrainedLM}. These models, primarily encompassing BERT-based architectures with bi-directional attention mechanisms \cite{devlinBERTPretrainingDeep2018} and GPT-based models utilizing auto-regressive learning schemas \cite{Radford2018ImprovingLU, brownLanguageModelsAre2020}, excel at building contextual representations and deriving nuanced understanding from extensive natural language data. As a result, complex language tasks such as document-level translation \cite{wang-etal-2023-document-level, zhuMultilingualMachineTranslation2024} and advanced question answering \cite{liLargeLanguageModel2024} can now be performed with minimal supervision.

When it comes to GEC area, LLMs excel at text manipulation tasks such as grammar correction \cite{10.1007/978-3-031-44699-3_7}, text rewriting \cite{shuRewriteLMInstructionTunedLarge2024}, and improving spoken language comprehension \cite{Chu2023QwenAudioAU, liUsingLargeLanguage2024}. By integrating generation and re-ranking, LLMs provide dynamic solutions for error correction and text refinement \cite{Yang2023GenerativeSR, Ma2023CanGL}. Models like ChatGPT \cite{Achiam2023GPT4TR} and ChatGLM \cite{zeng2023glmb} expand these capabilities further, performing text understanding, generation, and correction in unified frameworks. Despite their computational demands, smaller models like BART \cite{lewis2019Bart} and T5 \cite{10.5555/3455716.3455856} offer resource-efficient alternatives for specialized tasks, maintaining robust performance in constrained environments \cite{QIU2024200308}. 

By integrating LLM-based generative error correction, we can re-organize the slightly disordered output sequence of the neural spelling system into a more fluent language. This improves the overall system performance.

\section{Preliminaries}

In this section, we introduce the foundational concepts and notation used to describe our two-step neural processing system for EEG-based sentence generation. The system consists of two sequential modules: an encoder for EEG signal processing and a GenAI model to synthesize sentences.

\subsection{EEG Signal Encoder}
Let \(\mathbf{X} \in \mathbb{R}^{T \times C}\) represent the EEG input where \(T\) denotes the number of time steps and \(C\) the number of channels. The encoder function, denoted as \(f_{\text{enc}}\), transforms \(\mathbf{X}\) into a probability vector \(\mathbf{p} \in \mathbb{R}^{26}\), with each component \(p_i\) of \(\mathbf{p}\) representing the probability of the corresponding letter \(i\) from the English alphabet:
\begin{equation}
\mathbf{p} = f_{\text{enc}}(\mathbf{X}; \theta_{\text{enc}})
\end{equation}
where \(\theta_{\text{enc}}\) are the parameters of the encoder. The encoder utilizes a neural network architecture optimized for temporal and spatial features inherent in EEG data.

\subsection{GenAI Model for Synthesizing Sentences}
Upon generating the letter probabilities \(\mathbf{p}\), these are inputted into a GenAI model, \(g_{\text{trans}}\), which outputs a coherent sentence \(\mathbf{y}\). This model is based on a pretrained LLM that has been fine-tuned for the task of converting sequences of letter probabilities into grammatically correct and contextually relevant sentences:
\begin{equation}
\mathbf{y} = g_{\text{trans}}(\mathbf{p}; \theta_{\text{trans}})
\end{equation}
where \(\theta_{\text{trans}}\) represents the parameters of the translation model. The model leverages advanced natural language processing techniques to ensure semantic coherence and syntactic accuracy in the generated sentences.

These components are integrated to form a robust system for interpreting EEG signals and producing textual outputs, aiming to bridge the gap between neural activities and language expression.

\section{Methodology}

\subsection{Curriculum-based Neural Spelling Framework}

This section introduces the Curriculum-based Neural Spelling (CNS) Framework, a novel approach to enhancing the spelling accuracy from neural signals using a two-stage model. Initially, the model utilizes a Convolutional EEG Encoder to transform raw EEG data into a letter classification schema employing minimal neural samples and a tailored set of character categories. In the second stage, we integrate Curriculum Learning with large language models (LLMs) within a sequence-to-sequence framework to generate fluent sentences despite inherent noise and variability in EEG signals. This combination aims to leverage the strengths of advanced neural network architectures and sophisticated natural language processing techniques to overcome the challenges posed by low signal-to-noise ratios and the complex nature of EEG-based letter decoding. The architecture and its components are depicted in Fig.~\ref{fig:model_arch}, and this comprehensive setup promises significant advancements in brain-to-text communication technologies.

\subsubsection{Stage 1: Neural Letter Classifier}

\paragraph{Convolutional EEG Encoder}

The Convolutional EEG Encoder is designed as a two-layer Convolutional Neural Network (CNN), which is adept at processing EEG data for the purpose of embedding generation, as shown in Fig.~\ref{fig:model_arch}D. This encoder operates in conjunction with \revised{contrastive learning (ConL)} techniques to enhance the discriminative power of the embeddings. The trajectory data is processed by a pre-trained ResNet18 \cite{7780459} model, which serves as a trajectory encoder. The core of our learning strategy is based on minimizing the contrastive learning loss ($loss_{\revised{ConL}}$), which is calculated as follows:

The loss function is defined as:
\begin{equation}
    loss_{\revised{ConL}} =  1 - \cos(\theta_{EEG}, \theta_{Traj})^2
\end{equation}

where $\theta_{EEG}$ and $\theta_{Traj}$ represent the embedding vectors from EEG data and trajectory data, respectively.

\paragraph{Classification Head}

The letter classification component is an integral part of the CNS Framework. It includes a fully connected linear layer that maps its input to a 26-node output layer, where each node represents a letter of the English alphabet. After this. This arrangement enables the system to predict letters based on the processed EEG data. A softmax function is applied to the output layer to obtain a probability distribution over the alphabet, as shown in Fig.~\ref{fig:model_arch}E, which is critical for the system's accuracy in spelling prediction. The classifier utilizes the cross-entropy loss to measure the discrepancy between the predicted probabilities and the actual distribution of the target letters. The crossentrypy loss ($loss_{CE}$) function is defined as follows:

\begin{equation}
loss_{CE} = -\sum_{c=1}^{M} y_{o, c} \log(p_{o, c})
\end{equation}

where $M$ is the number of classes (letters), $y_{o, c}$ is a binary indicator (0 or 1) if class label $c$ is the correct classification for observation $o$, and $p_{o, c}$ is the predicted probability of observation $o$ being of class $c$. This loss function effectively guides the learning process by penalizing deviations from the true label distribution, thus enhancing the system's ability to generate accurate and reliable spelling predictions from EEG signals.

\revised{
In the training phase, the final loss ($loss_{total}$) is computed as a weighted combination of the cross-entropy loss ($loss_{CE}$) and the contrastive loss ($loss_{\revised{ConL}}$):
\begin{equation}
loss_{total} = 0.35 \times loss_{CE} + 0.65 \times loss_{\revised{ConL}}.
\end{equation}
The weighting coefficients are selected empirically to balance classification accuracy and representation regularization.
}

\subsubsection{Stage 2: Curriculum Learning for Language Model Integration in Fluent Sentence Generation}

The inherently low SNR of EEG signals combined with the rigorous demands of experimental protocols exacerbates the challenges of prediction biases and data scarcity in letter-level classification. These limitations significantly impede the capacity of classifiers to decode fluent sentences. To mitigate these challenges, we propose the integration of LLMs within a sequence-to-sequence (seq2seq) framework aimed at enhancing sentence fluency and readability.

Traditional NLP models often struggle with inputs characterized by significant letter-level noise due to their reliance on sub-word-level tokenization and training predominantly on well-curated texts. As depicted in Table \ref{tab:ChatGPTletter_denoising}, without a nuanced understanding of the behaviors specific to subject-based letter classifiers, models like ChatGPT4 can generate sentences that diverge significantly from the intended content, typically relying on generalized knowledge of language structure and common phrases.

\begin{table}[tp]
\centering
\caption{Challenges in Letter-Level Noise Denoising with ChatGPT 4}
\label{tab:ChatGPTletter_denoising}
\begin{tabular}{|p{0.9\linewidth}|}
\hline
\textbf{Ground Truth:} This civilization, known as the 'Ayar,' possesses an advanced understanding of nature and the power to manipulate the very fabric of reality. \\
\hline
\textbf{Prompt:} You are an expert on letter denoising of sentences. The letters from the following sentence are corrupted with noise. Can you guess and output the correct sentence? \{decoder\_output\} \\
\hline
\textbf{Input:} tvmg jijifizmhiog wnewn ms tve awmt pesowogeo in iqeanjqv onzerstmgvmng oz nmmxrw mgz txw peeer to mmgipxfamw mxe eefw zavfij ov feifatw \\
\hline
\textbf{Prediction:} Time is the most precious thing in life, understanding it makes the effort to simplify the world of survival or benefit. \\

\hline
\end{tabular}
\end{table}

\paragraph{Curriculum Learning Approach}

To address these constraints, we introduce a curriculum learning strategy tailored for pretrained LLMs. This strategy is designed to adapt the models to the latent distributions specific to subject-based handwriting patterns. By progressively increasing the relevance to the domain and the complexity of the tasks during training, our curriculum learning approach methodically fine-tunes LLMs. This targeted training enables the models to achieve better alignment with the idiosyncratic characteristics of subject-specific letter decoders, thereby enhancing their capacity to handle noisy inputs and improving overall sentence generation:

\begin{enumerate}
    \item \textbf{Initial Phase:} Start with tasks involving high-frequency, low-noise samples to establish baseline language structures.
    \item \textbf{Intermediate Phase:} Gradually introduce more complex and noisier data, increasing exposure to real-world variability.
    \item \textbf{Advanced Phase:} Focus on fine-tuning with the highest noise levels and the most challenging samples to ensure robustness and fluency under the most adverse conditions.
\end{enumerate}

This structured approach not only streamlines the adaptation process but also significantly boosts the model’s ability to generalize from noisy, imperfect inputs to coherent and contextually accurate text outputs.

\paragraph{Probabilistic Character Sampling from the Neural Letter Classifiers}

To evaluate the efficacy of our curriculum-based LLMs, we analyze the letter category distributions, $C_{i,\omega}$, for letter $\omega$ in sample $i$. These distributions are obtained from the softmax-normalized outputs of the classification head. We compute the aggregate Top-$K$ distributions, $C^*_{\omega}$, as follows:
\begin{equation}
    C^*_{\omega} = \frac{\sum_{i=1}^N C_{i,\omega}}{N_{\omega}}
\end{equation}
where $N_{\omega}$ denotes the count of samples for letter $\omega$. From this, the letter $\omega$ is then probabilistically sampled based on its presence within the Top-$K$ distribution. For instance, consider the ground truth (GT) sentence, "As he chases down rogue time travelers...". If the aggregated probability for 'A' at the start, $C^*_{A}$, is 0.2 and for 'R', $C^*_{R}$, is 0.6, the classifier might predict "Rs" in place of "As".

\paragraph{Implementation of Noise Adaptation}
\revised{
Noise adaptation refers to the process of fine-tuning the language model on EEG decoder outputs that contain systematic letter-level errors, enabling the model to recognize and correct typical noise patterns arising from neural decoding during inference.
}
In this phase, we incorporate the noisy predictions from the Neural Letter Classifier (NLC) into the LLM's training regimen. Specifically, we replace selected characters in the training texts with their corresponding noisy variants from the Top-$K$ output distributions. The LLM is then fine-tuned to de-noise this altered text. This structured noise adaptation process is designed to enhance the model's robustness, enabling it to effectively manage and correct noisy inputs during inference.

\paragraph{Curriculum Arrangement} In our curriculum learning method, training examples $\mathbf{D}$ are arranged according to a complexity score $\mathbf{C}$, in a multi-stage setting $\{T_i : i=1,2,\cdots, N\}$. In each stage $T_i$, a percentage $c\in\mathbf{C}$ of the letters from the input sentences will be permuted by the letter prediction distribution of the letter-classifier model, with a linear increase in the complexity score $c$.

\paragraph{Supervised Fine-tuning} We employ bi-directional attention during training for this seq2seq generation task, as shown in Fig.~\ref{fig:model_arch}(F). The fine-tuning objective combines the denoising training objective with the causal language modeling objective \cite{lewis2019Bart}, facilitating the model's use of previously generated tokens to guide the prediction of subsequent tokens, thereby reducing the influence of noisy letter-level inputs.
\begin{equation}
\mathcal{L}_{LMft} = - \sum_{t=1}^T \log P(y_t \mid \mathbf{x}, y_{<t})
\end{equation}
where $y_{<t}$ represents the previously generated tokens, and $\mathbf{x}$ is the noisy input sentence. In the subsequent Section \ref{sec:stage2result}, we will explore the performance benefits of the fine-tuned LLMs pretrained model utilizing our curriculum learning method in comparison to direct and non-fine-tuning approaches. Additionally, we will examine how variations in model size influence performance metrics.

\revised{The sensitivity analyses of the proposed curriculum learning (CL) strategy with respect to two key factors, noise injection strength and the number of curriculum stages, as well as the influence of different curriculum schedule shapes on model performance, are presented in Supplemental Sections I.A and I.B.}

\section{Experiments and Results}

\subsection{Participants}
We recruited thirty-two right-handed, healthy individuals (P1–P32), all native English speakers from Australia, comprising 16 males and 16 females with an average age of \( M_{\text{age}} = 24.94 \pm 0.29 \) years. Each participant had normal or corrected-to-normal vision and reported no history of mental health disorders. Four subjects were removed from the dataset due to misunderstanding of the experimental instructions. \revised{An a priori power analysis (G*Power v3.1.9.7) indicated that approximately 30 participants are sufficient to detect medium within-subject effects (Cohen’s \( d_z = 0.5 \)) at a significance level of \( \alpha = 0.05 \). The final sample size therefore meets standard statistical power requirements for within-subject EEG analyses.} Ethical approval was granted by the University of Technology Sydney's Ethics Committee (Approval No. UTS HREC REF: ETH23-8036). Informed consent was obtained in writing from each participant prior to the experiment.

\subsection{Instrumentation}
Handwriting movements and EEG signals were recorded simultaneously at sampling rates of 60 Hz and 1000 Hz, respectively. Handwriting trajectories were captured using a custom-developed application built on the PsychoPy platform, integrated with a Wacom Intuos Pro Medium tablet and Wacom Pro Pen 2 stylus featuring 8192 levels of pressure sensitivity. This application was designed to log critical events and features of each handwriting motion, including timestamps, \(x\)- and \(y\)-coordinates, rotation angles, force, and pen state codes for pen-down (contact with the surface), pen-move, and pen-up (lifting from the surface) for each digitalized point along the trajectory.

EEG signals were recorded using a 64-channel Neuroscan amplifier (Curry 9; Australia), with electrode placements following the 10–20 international system~\cite{jasper1958ten}. A ground electrode was referenced to maintain signal consistency.

Synchronization of handwriting trajectories with EEG signals was achieved by embedding time markers of key events within the EEG stream. Specifically, each first pen-down event of symbols written within each block on the tablet was marked for temporal alignment, as illustrated in Fig.~\ref{fig:model_arch}(B).

\subsection{Experimental Design}
The experiment was conducted in a sound-attenuated room. Participants were seated comfortably at a desk, positioned approximately 35 cm from a tablet, which was placed in an optimal location on the desk for each individual.

The task followed an event-related design, with extended breaks provided after every 10 trials. During these breaks, participants could rest for an unlimited duration, initiating the next trial by pressing the space key. Each trial, as illustrated in Fig.~\ref{fig:model_arch}A, began with a fixation cross displayed for 1000 ms with a random jitter of \(\pm\) 500 ms. Subsequently, a letter was presented for 2000 ms, cueing the character participants were to write in the handwriting task. During this phase, participants used a stylus to write the letter slowly within a 3000 ms time frame. A 2000 ms relaxation interval followed each handwriting task.

The sequence of the 26 letters was randomized for each participant, with each letter repeated 25 times, resulting in a total of 650 trials \((25 \times 26)\) per participant. The entire session lasted approximately 2 hours.

\begin{figure}[tp]
    \centering
    \includegraphics[width=1\linewidth]{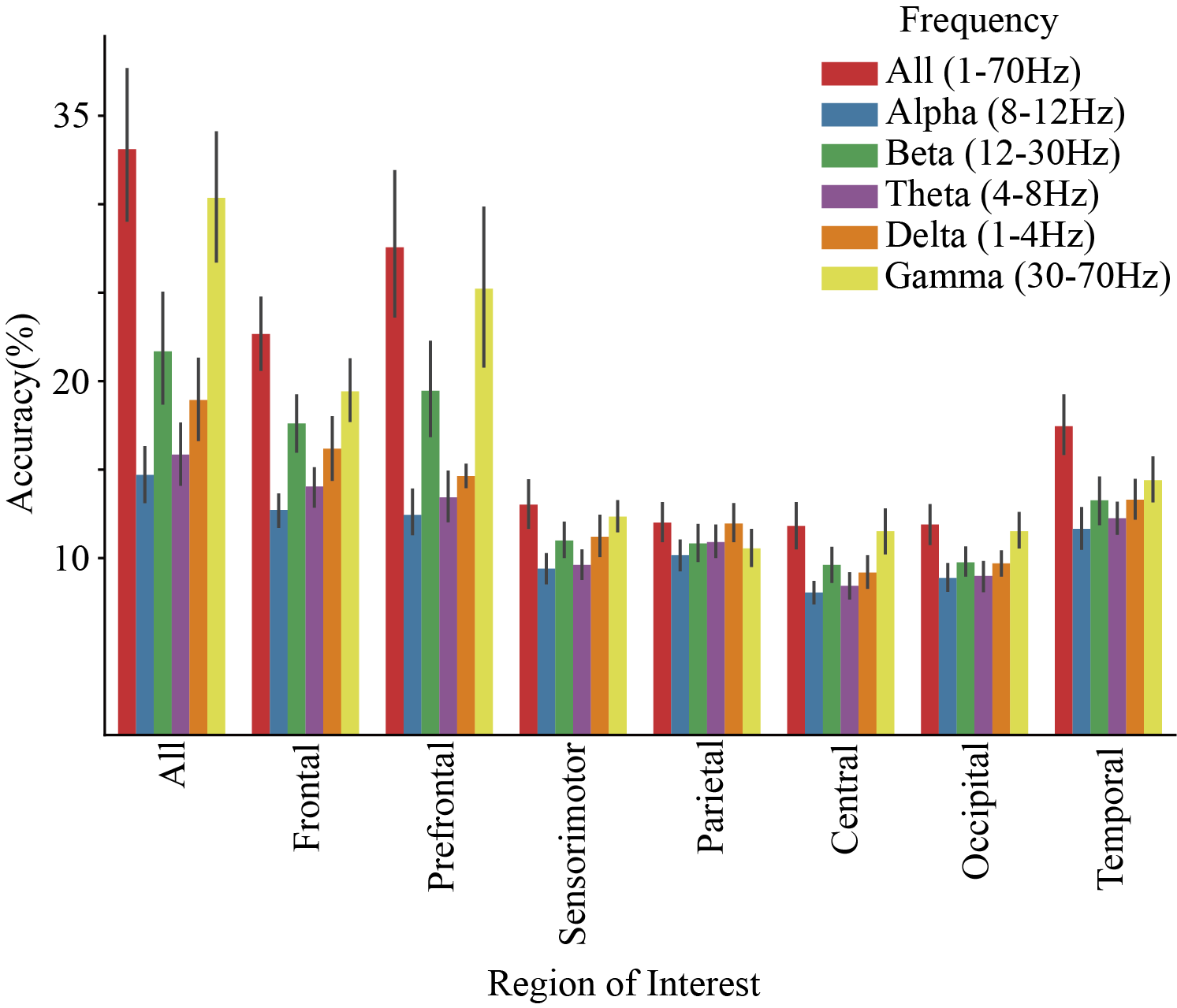}
    \caption{Top 1 Accuracy of CNN w/ \revised{ConL} Model Across Different ROIs and Frequency Bands.}
    \label{fig:roi_freq}
\end{figure}

\subsection{Data Processing}

\begin{figure}[b]
    \centering
    \includegraphics[width=1\linewidth]{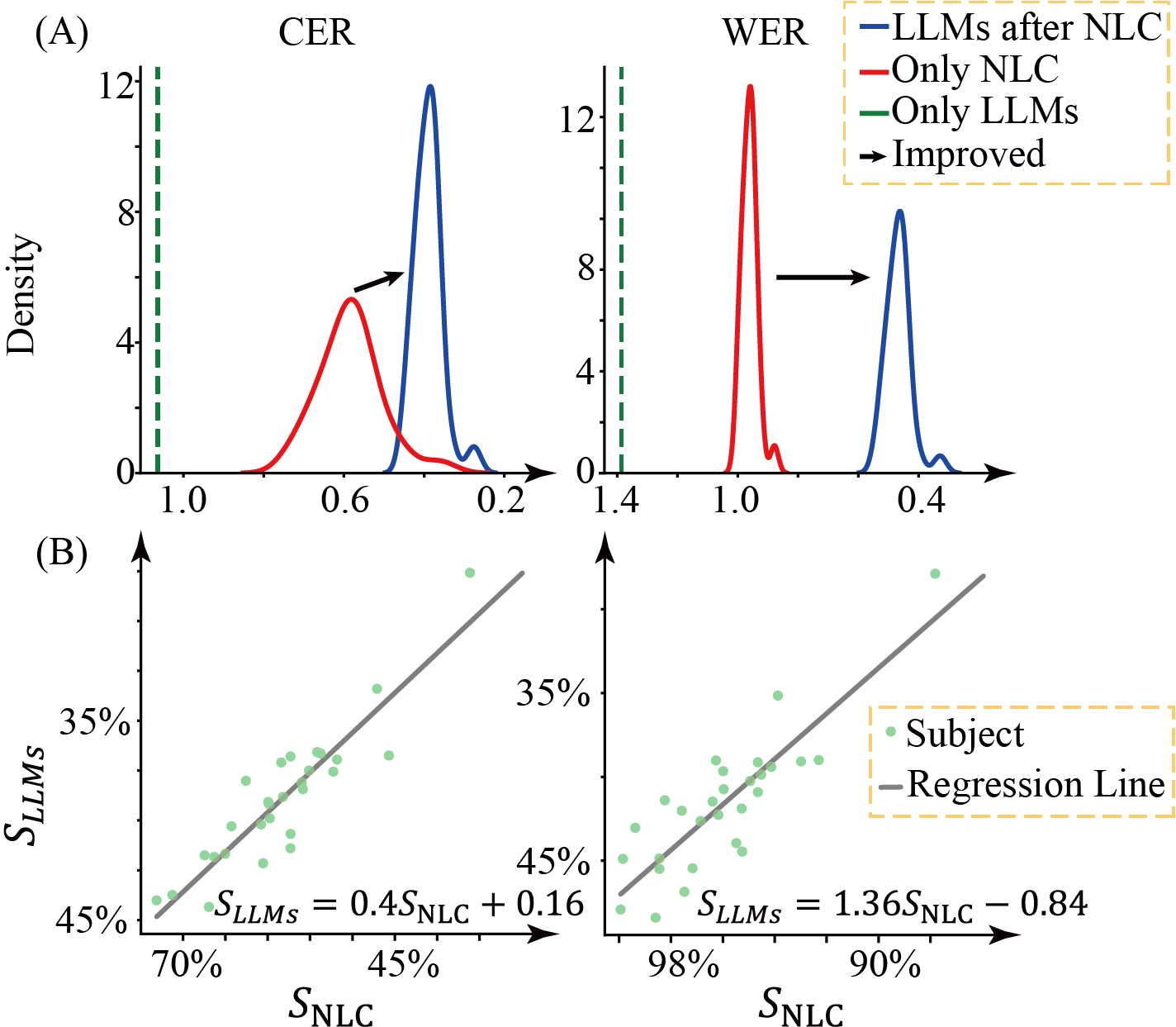}
    \caption{\revised{
    \textbf{(A):} The Kernel Density Estimate (KDE) distributions among only neural letter classifier (NLC), only LLMs (Large Language Models, the baseline), and LLMs after NLC.
    \textbf{(B):} The linear regression between the CER and WER scores of NLC ($S_{NLC}$) and LLMs ($S_{LLMs}$).}
    }
    \label{fig:EncodervsDecodercerwer}
\end{figure}

\paragraph{EEG Preprocessing}
EEG data preprocessing was performed using MNE (version 1.6.0)~\cite{Gramfort2013MEGAE} in Python 3.10.13. 
\revised{The pipeline began with a bandpass filter between 1 and 70~Hz to attenuate slow drifts and high-frequency noise, followed by notch filtering at 50, 100, and 150~Hz to suppress line noise and its harmonics.} 
\revised{EEG signals were then epoched from $-1$~s to $3$~s relative to the first pen-down event in each trial, capturing preparatory, execution, and post-execution neural activity.} 
\revised{Artifact handling was performed in two stages. First, automated channel-level artifact suppression was applied using the Autoreject framework~\cite{JAS2017417} to detect and interpolate noisy channels within each epoch; no trials were discarded at this stage. This step improves data quality for subsequent analysis while preserving the full trial structure.} 
\revised{Second, Independent Component Analysis (ICA) was applied to the epoched data after channel interpolation. ICA decomposition was performed using the FastICA algorithm implemented in MNE. Independent components were inspected following established EEG practice based on their temporal dynamics, spatial topographies, and spectral characteristics. Only clearly identifiable ocular artifacts (e.g., blink-related components and other clearly non-neural artifacts) were removed using a conservative manual procedure, with fewer than three components removed per subject.} 
\revised{Finally, the EEG data were re-referenced to the common average, and baseline correction was applied using the pre-event interval to ensure consistency across channels.}

\begin{table}[t]
\centering
\caption{\revised{Paired comparison between the proposed CNN-based model and representative baselines,
with and without contrastive learning (ConL). Accuracy is reported as mean $\pm$ standard deviation (\%).
Statistical significance is assessed using paired t-tests across subject-sessions,
with $^*$ $p<0.05$, $^{**}$ $p<0.01$, and $^{***}$ $p<0.001$, comparing each baseline against the proposed CNN model.}}
\label{tab:paired_ttest_models}
\resizebox{0.48\textwidth}{!}{

\revised{
\begin{tabular}{llccc}
\toprule
\textbf{Model} & \textbf{Setting} & \textbf{Top-1} & \textbf{Top-3} & \textbf{Top-5} \\
\midrule

\textbf{CNN (Proposed)}
& w/o ConL
& $27.23 \pm 12.10$
& $51.34 \pm 14.78$
& $65.29 \pm 12.08$ \\
& w/ ConL
& $33.10 \pm 11.54$
& $57.77 \pm 12.89$
& $71.46 \pm 11.11$ \\
\midrule

LSTM
& w/o ConL
& $18.22 \pm 3.46^{***}$
& $31.21 \pm 6.71^{***}$
& $41.66 \pm 6.42^{***}$ \\
& w/ ConL
& $25.74 \pm 3.51^{**}$
& $37.85 \pm 4.36^{***}$
& $49.00 \pm 6.07^{***}$ \\
\midrule

Transformer
& w/o ConL
& $15.54 \pm 4.40^{***}$
& $27.56 \pm 4.14^{***}$
& $36.44 \pm 4.02^{***}$ \\
& w/ ConL
& $23.99 \pm 3.14^{***}$
& $33.22 \pm 3.00^{***}$
& $43.92 \pm 3.50^{***}$ \\
\midrule

EEG-Deformer
& w/o ConL
& $17.94 \pm 4.46^{***}$
& $30.02 \pm 4.12^{***}$
& $38.82 \pm 4.04^{***}$ \\
& w/ ConL
& $28.99 \pm 3.14^{*}$
& $38.22 \pm 3.00^{***}$
& $48.92 \pm 3.50^{***}$ \\
\midrule

Conformer
& w/o ConL
& $14.59 \pm 4.57^{***}$
& $26.54 \pm 4.19^{***}$
& $35.28 \pm 3.73^{***}$ \\
& w/ ConL
& $25.56 \pm 3.29^{**}$
& $34.37 \pm 3.58^{***}$
& $45.54 \pm 3.22^{***}$ \\
\midrule

EEGFormer
& w/o ConL
& $15.21 \pm 4.97^{***}$
& $26.87 \pm 4.67^{***}$
& $35.74 \pm 3.74^{***}$ \\
& w/ ConL
& $26.07 \pm 3.38^{**}$
& $35.16 \pm 3.19^{***}$
& $46.46 \pm 3.86^{***}$ \\
\midrule

TCNet
& w/o ConL
& $16.34 \pm 4.63^{***}$
& $28.64 \pm 4.28^{***}$
& $37.46 \pm 3.96^{***}$ \\
& w/ ConL
& $27.53 \pm 3.17^{**}$
& $36.82 \pm 3.30^{***}$
& $47.17 \pm 3.60^{***}$ \\

\bottomrule
\end{tabular}
} 
}
\end{table}

\paragraph{EEG Feature Extraction}
Power Spectral Density (PSD) was extracted as the primary feature for model prediction. To compute PSD, an Fast Fourier Transform (FFT) was used capturing frequencies between 1 Hz and 70 Hz, as shown in Fig.~\ref{fig:model_arch}C. The PSD for a given signal \( x(t) \) was calculated using the formula:

\begin{equation}
\label{equ:psd}
\text{PSD}(f) = \frac{1}{N} \left| \sum_{t=0}^{N-1} x(t) e^{-i 2 \pi f t / N} \right|^2
\end{equation}

where \( f \) represents the frequency, and \( N \) is the total number of time points in the signal. This approach provided a detailed frequency profile of the EEG signals, essential for downstream analyses.

\paragraph{Trajectory Processing}
Trajectory data was processed by applying min-max normalization to the \( x(t) \) and \( y(t) \) coordinates, which were subsequently mapped onto a 28 x 28 pixel grid. To represent temporal progression, the intensity values along the trajectory were scaled from 50 to 255, illustrating the passage of time, as shown in Fig.~\ref{fig:model_arch}B. This approach allowed the temporal aspects of each handwriting movement to be visually represented in the spatial grid format.

\subsection{Stage 1: Neural Letter Classifier}

\begin{table*}[t]
\centering
\caption{Main Decoding Results}
\label{tab:decoder-main}
\begin{tabular}{l c cccc cccc cc}
\toprule
\multirow{2}{*}{Arch} & \multirow{2}{*}{Curriculum} 
& \multicolumn{4}{c}{BLEU (\%)} 
& \multicolumn{4}{c}{ROUGE (\%)} 
& \multicolumn{2}{c}{Error Rates (\%)} \\
\cmidrule(lr){3-6} \cmidrule(lr){7-10} \cmidrule(lr){11-12}
 &  & 1 & 2 & 3 & 4 & 1 & 2 & L & Lsum & CER & WER \\
\midrule
\multirow{2}{*}{Bart-base}
& no  & 25.74 & 10.77 & 5.02 & 2.68 & 22.78 & 3.80 & 18.51 & 18.50 & 92.61 & 120.32 \\
& yes & 27.55 & 11.46 & 5.64 & 3.17 & 24.07 & 4.04 & 19.51 & 19.50 & 89.63 & 113.22 \\
\midrule
\multirow{2}{*}{Bart-large}
& no  & 58.52 & 47.86 & 40.91 & 35.76 & 58.18 & 41.13 & 57.11 & 57.10 & 44.96 & 53.54 \\
& yes & 64.89 & 55.56 & 49.27 & 44.42 & 64.63 & 49.91 & 63.72 & 63.71 & 38.94 & 46.68 \\
\bottomrule
\end{tabular}
\end{table*}

\begin{table*}[t]
\centering
\caption{Ablation on the impact of space separation between words during neural spelling}
\label{tab:ablation-space}
\begin{tabular}{l l cccc cccc cc}
\toprule
\multirow{2}{*}{Arch} & \multirow{2}{*}{Space} 
& \multicolumn{4}{c}{BLEU (\%)} 
& \multicolumn{4}{c}{ROUGE (\%)} 
& \multicolumn{2}{c}{Error Rates (\%)} \\
\cmidrule(lr){3-6} \cmidrule(lr){7-10} \cmidrule(lr){11-12}
 &  & 1 & 2 & 3 & 4 & 1 & 2 & L & Lsum & CER & WER \\
\midrule
\multirow{2}{*}{Bart-base}
& no  & 40.14 & 25.47 & 18.24 & 13.97 & 35.24 & 15.12 & 32.00 & 31.97 & 66.46 & 83.10 \\
& yes & 45.46 & 32.01 & 24.92 & 20.37 & 41.15 & 21.96 & 38.30 & 38.27 & 60.89 & 76.22 \\
\midrule
\multirow{2}{*}{Bart-large}
& no  & 58.52 & 47.86 & 40.91 & 35.76 & 58.18 & 41.13 & 57.11 & 57.10 & 44.96 & 53.54 \\
& yes & 64.89 & 55.56 & 49.27 & 44.42 & 64.63 & 49.91 & 63.72 & 63.71 & 38.94 & 46.68 \\
\bottomrule
\end{tabular}
\end{table*}

\begin{table*}[ht]
\centering
\caption{Ablation Study on the Impact of Letter Classification Sampling Space}
\label{tab:ablation-letter-space}
\begin{tabular}{l c cccc cccc cc}
\toprule
\multirow{2}{*}{Model Arch} & \multirow{2}{*}{Sample Space}
& \multicolumn{4}{c}{BLEU (\%)} 
& \multicolumn{4}{c}{ROUGE (\%)} 
& \multicolumn{2}{c}{Error Rates (\%)} \\
\cmidrule(lr){3-6} \cmidrule(lr){7-10} \cmidrule(lr){11-12}
 &  & 1 & 2 & 3 & 4 & 1 & 2 & L & Lsum & CER & WER \\
\midrule
\multirow{7}{*}{Bart-base}
& 3  & 58.52 & 47.86 & 40.91 & 35.76 & 58.18 & 41.13 & 57.11 & 57.10 & 44.96 & 53.54 \\
& 7  & 51.79 & 39.32 & 31.89 & 26.80 & 49.31 & 30.32 & 47.60 & 47.60 & 53.69 & 64.31 \\
& 11 & 45.98 & 32.34 & 24.72 & 19.83 & 42.78 & 22.54 & 40.56 & 40.54 & 60.31 & 72.27 \\
& 15 & 44.62 & 30.68 & 23.06 & 18.29 & 40.97 & 20.65 & 38.63 & 38.61 & 62.08 & 74.78 \\
& 19 & 43.76 & 29.65 & 22.10 & 17.42 & 40.01 & 19.59 & 37.53 & 37.52 & 63.12 & 76.09 \\
& 23 & 43.48 & 29.31 & 21.74 & 17.07 & 39.65 & 19.22 & 37.11 & 37.09 & 63.46 & 76.55 \\
& 26 & 48.14 & 34.61 & 27.03 & 22.10 & 44.44 & 24.68 & 42.14 & 42.14 & 58.78 & 71.18 \\
\midrule
\multirow{7}{*}{Bart-large}
& 3  & 64.89 & 55.56 & 49.27 & 44.42 & 64.63 & 49.91 & 63.72 & 63.71 & 38.94 & 46.68 \\
& 7  & 55.22 & 43.30 & 35.94 & 30.73 & 52.74 & 34.52 & 51.12 & 51.13 & 50.72 & 60.95 \\
& 11 & 51.25 & 38.41 & 30.86 & 25.76 & 48.06 & 28.87 & 46.06 & 46.06 & 55.41 & 66.70 \\
& 15 & 49.28 & 36.03 & 28.44 & 23.44 & 45.73 & 26.16 & 43.52 & 43.53 & 57.64 & 69.62 \\
& 19 & 48.44 & 35.01 & 27.43 & 22.48 & 44.94 & 25.19 & 42.67 & 42.67 & 58.31 & 70.47 \\
& 23 & 47.94 & 34.37 & 26.78 & 21.85 & 44.30 & 24.48 & 41.98 & 41.97 & 59.13 & 71.60 \\
& 26 & 48.13 & 34.66 & 27.11 & 22.20 & 44.44 & 24.74 & 42.12 & 42.13 & 58.92 & 71.36 \\
\bottomrule
\end{tabular}
\end{table*}

\subsubsection{Model Performance Comparison}

\revised{We evaluated the performance of different neural network architectures for EEG-based letter decoding, including LSTM, transformer-based, EEG-Deformer~\cite{EEGDeformer}, Conformer~\cite{EEGConformer}, EEGFormer~\cite{EEGformer}, and TCNet~\cite{eegtcnet}, with and without \revised{ConL}. The comparison focuses on Top-1, Top-3, and Top-5 classification accuracy and is summarized in Table~\ref{tab:paired_ttest_models}.

Across all evaluated models, the proposed CNN-based architecture consistently achieved the highest decoding accuracy under the current experimental setting. In particular, the CNN with ConL attained a Top-1 accuracy of $33.10\% \pm 11.54\%$, a Top-3 accuracy of $57.77\% \pm 12.89\%$, and a Top-5 accuracy of $71.46\% \pm 11.11\%$, outperforming all recurrent and transformer-based baselines. Paired t-tests across subject-sessions indicate that these improvements are statistically significant in most comparisons, with multiple-comparison correction applied using the Benjamini–Hochberg procedure.
}

\begin{figure*}
    \centering
    \includegraphics[width=0.8\linewidth]{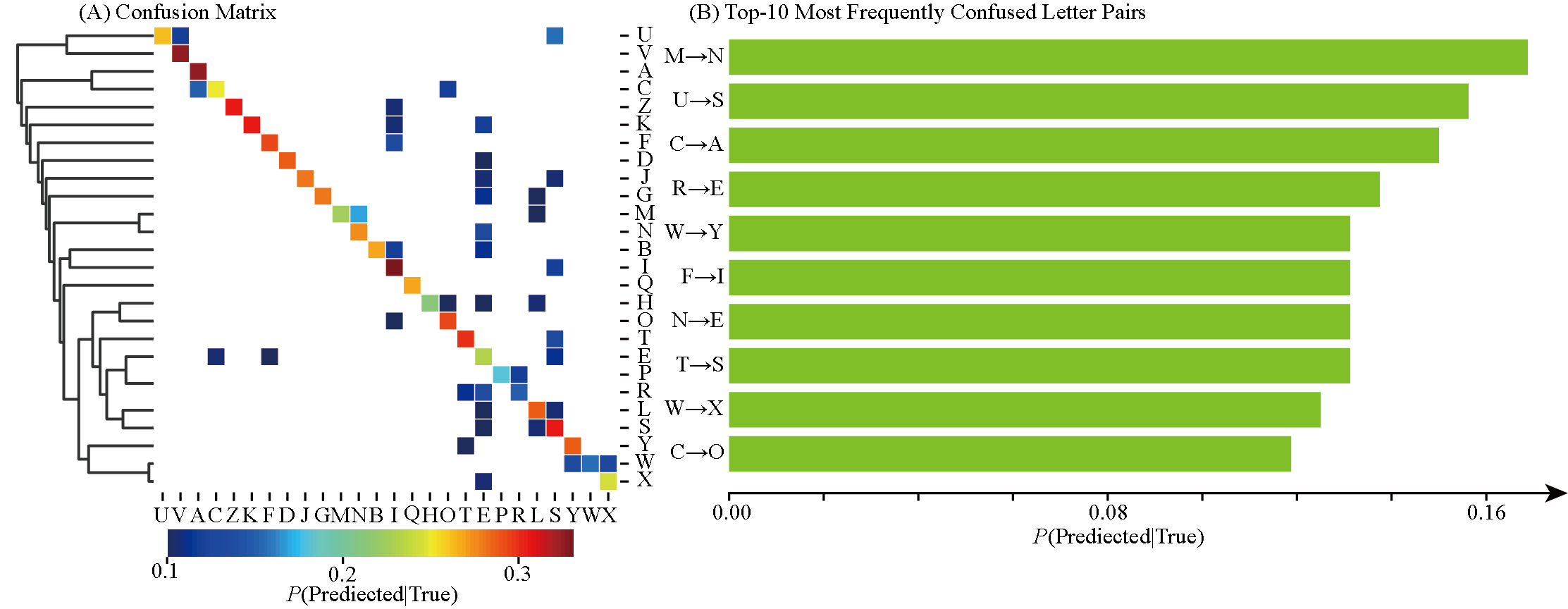}
    \caption{\revised{
    \textbf{(A):} Confusion matrix of letter decoding performance with hierarchical clustering. The matrix shows the averaged conditional probability $P(\text{predicted}\mid\text{true})$ (colored $P>0.1$) for 26 handwritten letters, reordered using hierarchical clustering based on confusion profiles.
    \textbf{(B):} Top-10 Most Frequently Confused Letter Pairs.}
    }
    \label{fig:ConfusionMatrix}
\end{figure*}

\revised{
\paragraph{Confusion matrix analysis.}
Fig.~\ref{fig:ConfusionMatrix}(A) illustrates the confusion matrix of the averaged conditional probability
$P(\text{predicted}\mid\text{true})$ for 26 handwritten letters. Overall decoding performance varies
substantially across letters. Letter \emph{I} achieves the highest classification accuracy (33.12\%),
whereas letter \emph{R} exhibits the lowest accuracy (15.00\%), indicating pronounced letter-specific
decoding difficulty.

Certain letters show consistently higher confusion rates. In particular, letter \emph{E} demonstrates
relatively poor performance (23.13\%) and is frequently misclassified as visually or motorically similar
letters, including \emph{D, J, G, N, B, H, R, L, S,} and \emph{X}. This widespread confusion suggests that
\emph{E} shares common stroke components and motor primitives with multiple letter classes, making it
difficult to disentangle under non-invasive EEG constraints.

Fig.~\ref{fig:ConfusionMatrix}(B) further summarizes the most frequently confused letter pairs. Among them,
the \emph{M--N} pair exhibits the highest mutual confusion probability, which can be attributed to their
highly similar handwriting trajectories characterized by repeated vertical strokes and comparable
temporal motor patterns. Overall, these results indicate that misclassifications are highly structured
rather than random, reflecting intrinsic similarities in neural and motor representations rather than
decoder instability.

}

\subsubsection{Model Performance Across Different ROIs and Frequencies}

This section investigates the impact of different ROIs and frequency bands on the Top 1 accuracy of a CNN model augmented with \revised{ConL}. The analysis focuses on the PSD extracted from specific bands and ROIs to determine their relative importance in performance metrics.

Fig.~\ref{fig:roi_freq} illustrates the results of this comparison. Utilizing all features from all bands and ROIs yields the highest Top 1 accuracy, indicating the advantage of a comprehensive feature set. The Gamma band stands out among the frequency bands, showing the largest performance improvement, underscoring its significance in EEG-based models. Regarding the ROIs, the PFC achieves the first-highest Top 1 accuracy, followed by the whole frontal cortex. Subsequent analyses show that the temporal and sensorimotor cortices rank lower, yet they contribute to overall model accuracy.

\revised{
\textbf{Informative EEG Feature Analysis.}
To explicitly investigate which EEG features are most informative for letter decoding, we analyzed decoding performance across both spectral and spatial dimensions. As shown in Fig.~\ref{fig:roi_freq}, frequency-band analysis reveals that high-frequency components, particularly the gamma band (30--70 Hz), contribute most strongly to Top-1 decoding accuracy. In contrast, lower-frequency bands (delta and theta) show relatively limited discriminative power for letter classification. Spatially, electrodes over the prefrontal and frontal cortices yield the highest decoding performance, followed by temporal and sensorimotor regions. Using all ROIs jointly achieves the best overall accuracy, indicating that neural letter representations are distributed across multiple cortical areas rather than localized to a single region. Together, these results suggest that informative features for non-invasive neural spelling are largely associated with high-frequency spectral activity in frontal--prefrontal networks associated with motor planning and language-related processing.
}

\subsection{Stage 2: Curriculum-based Large Language Model}
\label{sec:stage2result}

\subsubsection{Sentence Dataset}

For Stage 2 verification, we utilize the \textit{Story-Plots-1.3k} dataset from QuasarResearch, an expansion of the \textit{Neural-Story-v1} from NeuralNovel. This dataset is specifically designed for applications in creative text generation and narrative analysis, comprising 1,320 unique story plots. All stories are included in our testing set. The dataset is hosted on Hugging Face, available at \url{https://huggingface.co/datasets/QuasarResearch/story-plots-1.3k}.

\subsubsection{Sentence Decoding Performance}

Table \ref{tab:decoder-main} presents the results of sentence decoding from the encoder, utilizing the Top-k predicted distribution as input for the LLM decoder (where $k=3$). We evaluate the decoding performance across two different model sizes of the BART model \cite{lewis2019Bart}, namely `large` and `base`, with the trainable parameters equal to 406M and 140M, respectively. For both model sizes, our proposed curriculum learning method significantly enhances decoding performance across all evaluated metrics. Specifically, using the pretrained BART-large model, our method achieves a BLEU-4 score of $44.42\% \pm 2.2\%$ and a ROUGE-L score of $63.71\%\pm 0.86\%$. 

Additionally, Fig.~\ref{fig:EncodervsDecodercerwer}(A) compares the performances among only neural letter classifier, only LLMs (baseline), and LLMs after neural letter classifier, show the LLMs can improve the performance for this spelling-based language decoding. This observation underscores the potential of LLMs to enhance the language decoding task, suggesting that improvements in letter classification directly contribute to better sentence reconstruction outcomes.

Fig.~\ref{fig:EncodervsDecodercerwer}(B) revealing a positive linear relationship between the decoder's performance on the sentence decoding task and the neural letter classifier's performance. It shows that LLMs after neural letter classifier can significantly improve in word level.

\subsubsection{Visualization of Sentence Decoding Results}

For qualitative analysis, we visualize the decoding results of our method in \revised{Supplemental Table I}. The input samples were taken from subject P11, who exhibited the highest letter-level classification performance in our dataset. Notably, some phrases from the initial neural decoding are already partially informative and resemble the target sentence structure. For example, the fragment `rn thp unltrgfving...' bears clear similarity to the ground-truth phrase `in the unforgiving...'.

\revised{The LLM decoder then integrates these noisy letter-level predictions with learned word-transition statistics and linguistic constraints, enabling improved grammatical structure and word-level coherence. Importantly, the resulting sentences are not intended to be exact verbatim reconstructions of the ground truth. Instead, they represent semantically plausible and communicatively usable outputs that balance neural evidence with strong language priors.}

\revised{As illustrated in \revised{Supplemental Table I}, this process may introduce creative variations, particularly when the neural input is highly uncertain, reflecting a trade-off between symbol-level fidelity and language-level coherence. These qualitative results demonstrate the effectiveness of the proposed framework in refining linguistic outputs under noisy neural decoding conditions, while preserving consistency with the underlying neural evidence.}
%P11 -> 003-003

\begin{figure}[t]
    \centering
    \includegraphics[width=0.8\linewidth]{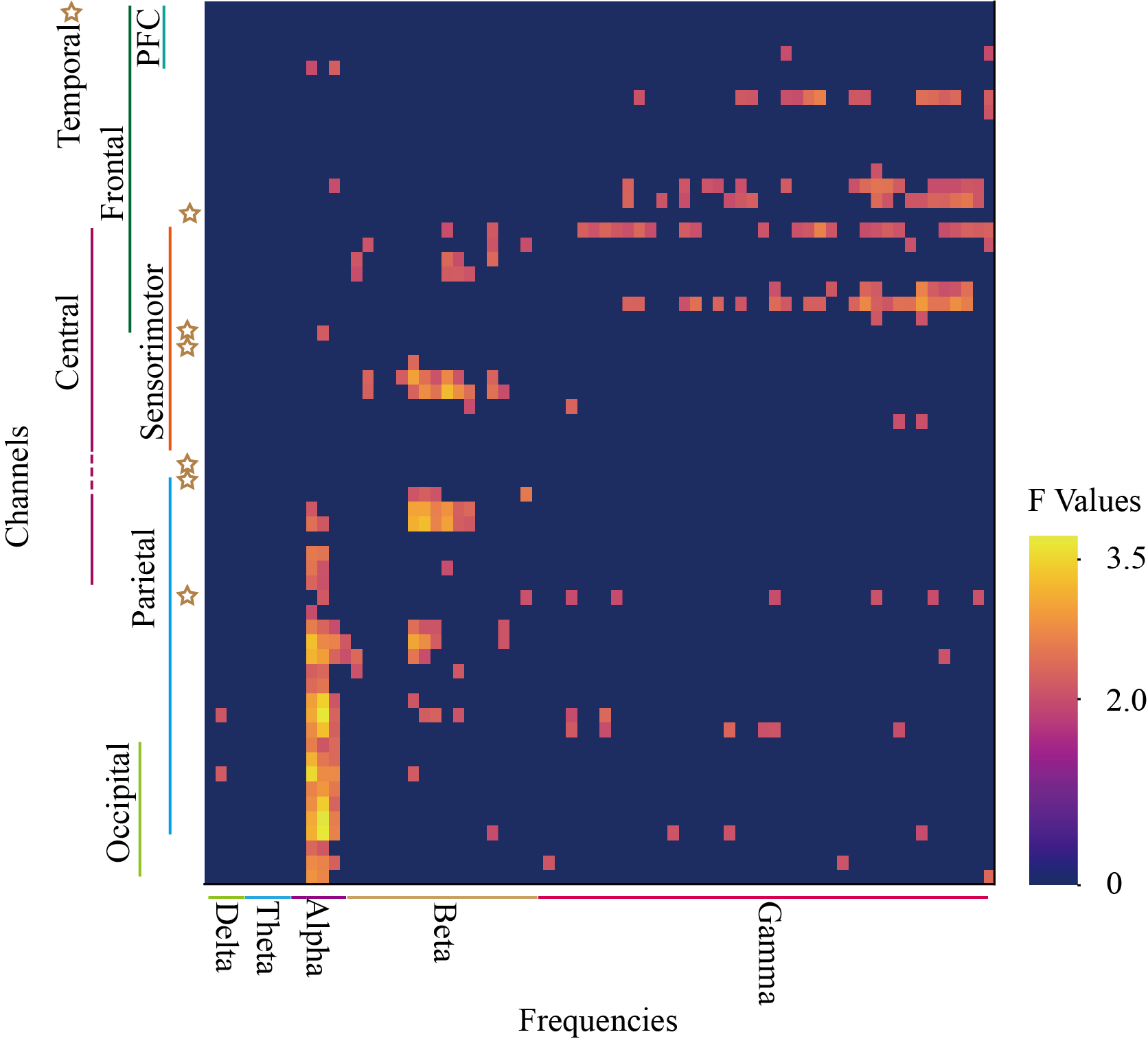}
    \caption{Significance of EEG spectral variations across different regions and frequencies, analyzed using F-statistics (F(25,675)) and corrected for multiple comparisons with the FDR\_BH approach (colors indicate $\textit{p}<0.05$).}
    \label{fig:F_values}
\end{figure}

\subsection{Ablation Study}

% \subsubsection{Ablation on Sentence Decoding}
% In this section, we conduct an ablation study to assess the impact of the input noise space on the model's performance. In the main experiment section of the decoder, we predominantly reported results from the top-3 classification distribution. Here, we extend our analysis to examine the robustness of the proposed model against the full noise space, encompassing all 26 letters. The results of this study are displayed in Table \ref{tab:ablation-letter-space}. 

% We observe that sampling letters from a broader prediction range leads to a decrease in performance. This decline is attributed to the increased difficulty for the decoder in learning subject-dependent transition patterns, thus complicating the task of accurately reconstructing the intended text.

\subsubsection{Ablation on the effect of letter sampling range}
In our analysis, we observe a decline in performance across all metrics when sampling from a larger prediction range. This outcome illustrates a trade-off between letter-level prediction accuracy and input diversity. To balance these factors effectively, we opt to utilize the top-3 prediction results for each letter to sample the letters spelled by the user, as detailed in Table \ref{tab:ablation-letter-space}. This approach optimizes our model's accuracy while maintaining a reasonable level of diversity in the input data.

\subsubsection{Ablation on the Effect of Space Between Words During Neural Spelling}

Results are presented in Table \ref{tab:ablation-space}. In conclusion, while incorporating spaces between words introduces an additional step for the user, it aids the model in more effectively denoising the input. However, the overall results demonstrate that our model is capable of denoising neural spelling text effectively even when spaces are not included in the input. This underscores the robustness of our approach in handling continuous streams of characters without explicit segmentation.

\section{Discussion}

This study presents a Curriculum-based Neural Spelling Framework comprising a Convolutional EEG Encoder and a \revised{ConL} module, demonstrating the feasibility of recognizing handwritten letters using non-invasive technologies. Furthermore, the system highlights the potential for real-world applications by integrating LLMs to generate free-form sentences based on spelling-based designs.

\begin{figure}[b]
    \centering
    \includegraphics[width=0.7\linewidth]{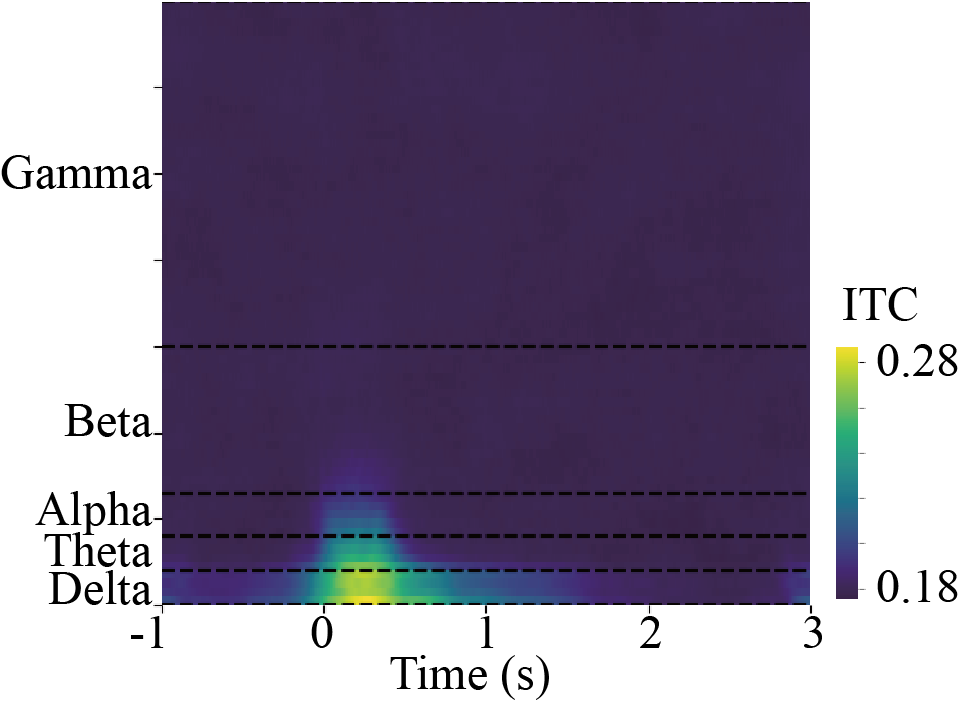}
    \caption{Visualization of ITC across different frequencies (1Hz to 70 Hz) during handwriting tasks. The color scale indicates the level of coherence, with warmer colors representing higher coherence.}
    \label{fig:itc}
\end{figure}

\subsection{Neurophysiological Correlates of Letter Recognition}

\begin{figure}[]
    \centering
    \includegraphics[width=1\linewidth]{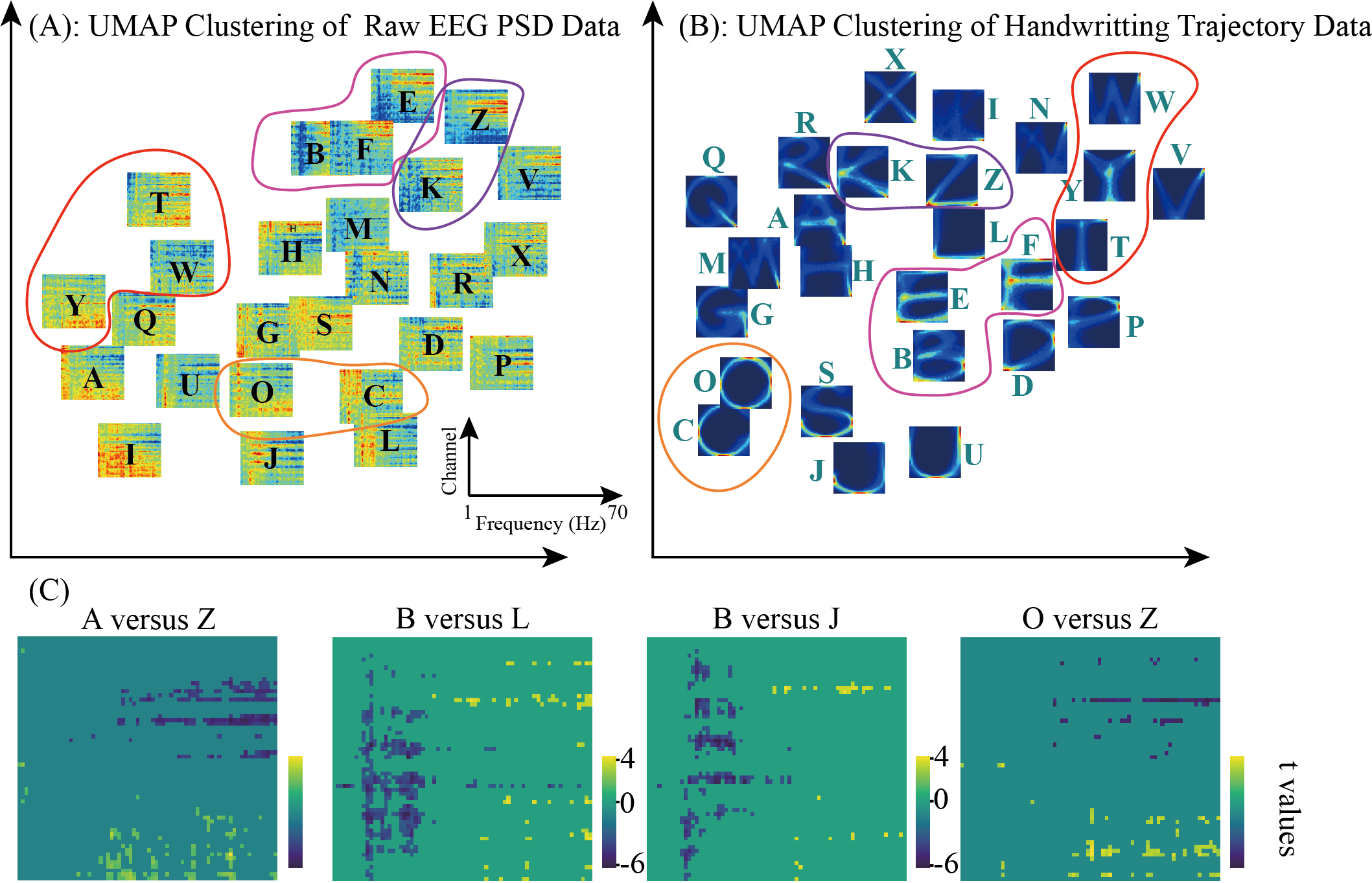}
    \caption{UMAP visualization of letter-specific neural representations. 
    \textbf{(A):} EEG PSD features with averaged baseline removed to enhance visualization clarity. 
    \textbf{(B):} Handwriting trajectories showing distinct clusters for letters TWY, BFE, ZK, and OC, which suggest similar neurocognitive processing pathways and minor differences in movement patterns. 
    \textbf{(C):} Significant differences between letter pairs A and Z, B and L, B and J, and O and Z, assessed via paired t-tests and corrected for multiple comparisons using the FDR\_BH method (colors indicate $\textit{p}<0.05$), highlighting the neurophysiological distinctions in letter processing.
    \revised{UMAP embeddings were computed and visualized on a per-subject basis to account for inter-individual EEG variability.}
}
    \label{fig:umap}
\end{figure}

\revised{
From a feature perspective, our results indicate that letter decoding is strongly associated with spectral-spatial EEG patterns rather than transient time-domain signals. In particular, gamma-band activity over frontal and prefrontal regions exhibits strong discriminative power across letters. This pattern is consistent with prior literature linking gamma activity to fine motor execution and higher-order cognitive processing.

In contrast, alpha-band activity shows letter-dependent modulation mainly in parietal and central regions, which is consistent with visuomotor integration and trajectory planning during handwriting. Lower-frequency oscillations, including delta and theta bands, demonstrate increased inter-trial coherence during movement initiation but contribute less directly to class separability. These findings are consistent with a potential functional differentiation in which high-frequency spectral features are more strongly associated with letter-specific decoding performance, while low-frequency dynamics appear more related to task coordination and motor timing.

The qualitative analysis of the Event-Related Potential (ERP) results is presented in Supplemental Section II.A.
}

Handwriting involves complex neural coordination, engaging multiple brain regions. \revised{Fig.} ~\ref{fig:umap} illustrates how specific letters correspond to neural activation (A) and handwriting trajectory (B). Some letters, such as TWY, BFE, ZK, and OC, show similar patterns in both neural and trajectory patterns. The small distance of these letters, suggests a shared representation in neural processing spaces.

Significant neurophysiological variations are highlighted in Fig.~\ref{fig:F_values}, where the F-statistics reveal prominent differences across cerebral regions at specific frequency bands. In particular, the alpha band shows substantial activity in the occipital and parietal lobes ($\textit{p} < 0.05$), underscoring their roles in visual processing and spatial integration~\cite{GALETTA2014626, Gottlieb2010SpatialAN}, essential for interpreting letter shapes and trajectories.

As demonstrated in Fig. \ref{fig:umap}(C), the neural pattern differences between letters exhibit complex and diverse styles across different brain regions and frequency bands ($\textit{p} < 0.05$). Specifically, the differences between letters A and Z are predominantly observed in the Gamma band within the Frontal and Parietal cortices ($\textit{p} < 0.05$). In contrast, the patterns for B and L are distinctly marked by Alpha band activity in the Parietal and Central cortices ($\textit{p} < 0.05$). Interestingly, the neural patterns for B and J are similar to those of B and L but feature less Gamma band activity in the Frontal cortex ($\textit{p} < 0.05$). Moreover, the patterns for O and Z resemble those between A and Z, albeit with a reduced difference in Gamma activity ($\textit{p} < 0.05$).  These observations highlight the unique neurophysiological pathways for each letter pair, reflecting varied motor and cognitive demands, and suggesting differences in cognitive processing levels.

The frontal cortex demonstrates increased gamma-band activity, reflecting its integrative function across sensory modalities and its pivotal role in higher cognitive processes, such as memory and decision-making~\cite{Kanayama2007CrossmodalEW, Schneider2008EnhancedEG, Roux12411, CHO20151332}, which are crucial for language processing and handwriting execution.

Moreover, the prefrontal cortex (PFC) emerges as a critical node, facilitating the integration of multi-modal information and mediating complex cognitive functions including executive control and working memory~\cite{Menon2021PFCnetworks, https://doi.org/10.1155/2017/1695290}. Similarly, temporal cortex activations are closely tied to language processing, with gamma activity playing a significant role in the neural decoding of language elements~\cite{willettHighperformanceSpeechNeuroprosthesis2023, anumanchipalliSpeechSynthesisNeural2019}.

Additionally, the sensorimotor cortex's involvement aligns with its role in governing motor control and sensory processing, fundamental to handwriting~\cite{SESSLE20163}. The broad gamma activity across these regions suggests a high-level synchronization of cortical activity, facilitating coherent cognitive representations necessary for complex tasks like letter recognition and differentiation~\cite{Misselhorn2019, Doesburg2008LargescaleGP, Bosman2014FunctionsOG}.

This study underscores the nuanced interplay of multiple brain regions during letter recognition and handwriting, paralleling findings from related fMRI studies~\cite{huthNaturalSpeechReveals2016}. The identified patterns across the frontal, parietal, and temporal cortices substantiate their critical roles in the cognitive and perceptual foundations of language and handwriting tasks. Utilizing all ROIs and frequency bands yields the best performance, as demonstrated in Fig. \ref{fig:roi_freq}.

\revised{
We emphasize that all neurophysiological interpretations are based on correlational analyses of sensor-level EEG signals. Accordingly, observed spectral and spatial differences are interpreted as task-dependent neural associations rather than direct evidence of causal mechanisms or localized neural representations.
}

\subsection{Inter-Trial Coherence in Delta Band During Handwriting Tasks}

Inter-Trial Coherence (ITC) measures phase synchronization across EEG trials, providing insights into consistent neural responses and cortical synchrony during event-related cognitive tasks \cite{makeig2004mining}

\begin{equation}
    ITC(f) = \left| \frac{1}{N} \sum_{k=1}^{N} e^{i \phi_k(f)} \right|
\end{equation}
where \( N \) is the number of trials, \( f \) represents the frequency, and \( \phi_k(f) \) is the instantaneous phase of the \( k \)-th trial at frequency \( f \). 

Our analysis, during handwriting tasks, particularly underscores the Delta frequency band, which exhibits significant coherence during the initial phase of action execution (0 to 1 second), as depicted in Fig.~\ref{fig:itc}. This heightened coherence suggests a robust engagement of neural circuits associated with low-frequency oscillations, crucial for the timing and coordination of motor movements.

Studies have indicated the involvement of Delta and Theta bands in motor tasks and cognitive processing \cite{lin2022effects}. For instance, increased ITC in these bands has been associated with task switching and inhibitory control in BCI paradigms \cite{Barios2017DeltaThetaIP}. Similar patterns of coherence in Delta-Theta ranges have been observed during motor performance improvements in stroke rehabilitation, reflecting neural reorganization and recovery \cite{Gyulai2021InterTC}. In addition, heightened Theta synchronisation has also been shown to correlate with navigation workload during physical navigation, supporting the role of low-frequency oscillations in coordinating cognitive–motor demands in ecologically valid settings \cite{do2020estimating, do2021retrosplenial}. The periodic auditory stimulation research shows a synchronization peak at 2 Hz in the Delta band, linking sensory-motor time coupling, which may parallel the neural dynamics observed during the rhythmic movements in handwriting \cite{Will2007BrainWS}.

Moreover, the Alpha band, while showing lower coherence compared to the Delta band during the initial phase of handwriting, plays a significant role in broader cognitive and motor control contexts. Increased Alpha ITC has been associated with enhanced visuomotor performance and is particularly evident in tasks requiring visual coordination, such as visuomotor tracking~\cite{Rilk2011AlphaCP}. Furthermore, variability in Alpha ITC has been linked to cognitive control efficiency within the frontoparietal network~\cite{Aaron2015VariabilityII}, an aspect crucial for complex task execution.

This differential engagement of frequency bands underscores the complexity of brain dynamics during fine motor tasks such as handwriting. While Delta, Theta, and Alpha oscillations may not yield the greatest contributions in NLC applications, their primary roles in initiating and coordinating motor output, coupled with their subsequent involvement in cognitive processing, illustrate a layered neural architecture that supports both the execution and cognitive integration of handwriting tasks. Although these bands do not dominate the neural decoding efforts, their influence is nonetheless crucial for the holistic understanding of motor control mechanisms.

% \subsection{The Learned distribution of GenAI Model}

% \subsection{Generative Capacity from English to Chinese Translation}
% In this section, we explore the adaptability of the second-stage GenAI model by transitioning from GEC English LLMs to Pinyin Input LLMs. As demonstrated in Table~\ref{tab:xxx}, the results indicate that the model maintains effective performance even when directly modified from English to Chinese, without the need for re-tuning the NLC. This suggests the generative capacity of this CNS framework for neural language decoding, powered by the outside extra lexicon and corpus.

\section{Conclusions}

This study introduces a CNS that leverages the advanced capabilities of GenAI to enhance spell-based neural language decoding tasks. Our approach is distinct in integrating a CNN with a curriculum-driven LLM, promoting an innovative hybrid method in the domain of BCIs. The framework's effectiveness is demonstrated through its application to EEG-based handwriting of all 26 letters, a novel endeavor in the field. The CNS framework notably achieves exemplary top-k accuracy across all subjects, underscoring the robustness of the EEG encoding model. Furthermore, our curriculum supervised fine-tuning method significantly advances the state of the art by enabling the LLM to effectively learn subject-specific letter transition patterns. This methodological innovation not only enhances sentence synthesis quality but also establishes a system-level strategy for compensating low-SNR neural decoding through language-level error correction.
\revised{This work addresses a translational feasibility question by demonstrating how full-alphabet neural spelling and curriculum-adapted generative language models can be integrated within non-invasive EEG constraints.} By seamlessly merging non-invasive EEG with GenAI, this study advances the methodological foundation of neural spelling systems and sets the stage for future investigations into more constrained paradigms, including online operation and reduced-movement or imagery-based settings.

\section{Limitations}
\revised{
The present study has several limitations. First, all experiments are conducted in an offline decoding setting, and system performance under real-time operation remains to be validated. Second, the current framework primarily relies on within-subject training, which limits cross-subject generalizability and scalability. Third, the dataset size per subject is relatively limited, constraining the training of larger or more expressive neural encoders. 

In addition, the proposed neural handwriting paradigm involves overt motor execution, which restricts direct applicability to individuals with severe motor impairments. We therefore position this paradigm as a high-SNR setting for studying full-alphabet neural spelling and language-level correction.
}

\section{Future Work}

\revised{
Future work will extend the proposed framework toward realistic online BCI deployment. Low-latency neural decoding can be achieved using lightweight convolutional models, while language-level correction can operate asynchronously, enabling responsive closed-loop interaction. Future studies will focus on end-to-end online validation, latency analysis, and long-term stability under continuous use, as well as extending the paradigm to reduced-movement or imagery-based settings to improve accessibility. In terms of scalability, future work will prioritize reduced-calibration and few-shot adaptation strategies rather than fully subject-independent decoding, which remains challenging for non-invasive EEG due to substantial inter-subject variability. Future work will also investigate constrained and confidence-aware generative decoding strategies to better balance expressive language reconstruction with reliability and user intent preservation in neural spelling systems.}
\revised{
Future work will explicitly address safety and hallucination risks in language-level decoding by incorporating confidence-aware constraints, task-specific vocabularies, and optional human-in-the-loop verification to ensure reliable and intent-preserving communication in clinical settings.
}
\revised{Finally, future work will systematically examine behavioral and task-related factors underlying inter-subject variability, including handwriting kinematics (e.g., speed and trajectory complexity) and effective trial counts after artifact rejection, to better contextualize individual differences in decoding performance.}

\begin{small}
\bibliographystyle{IEEEtran}
\bibliography{ref}
\end{small}

\end{document}